# A Study into patient similarity through representation learning from medical records

Hoda Memarzadeh

Department of Electrical and Computer Engineering, Isfahan University of Technology, Isfahan 84156-83111, Iran. Electronic address: `h.memarzadeh@ec.iut.ac.ir`

Nasser Ghadiri

Department of Electrical and Computer Engineering, Isfahan University of Technology, Isfahan 84156-83111, Iran. Electronic address: `nghadiri@iut.ac.ir`

Matthias Samwald

Institute for Artificial Intelligence, Center for Medical Statistics, Informatics, and Intelligent Systems, Medical University of Vienna, Vienna, Austria. Electronic address: `matthias.samwald@meduniwien.ac.at`

Maryam Lotfi Shahreza

Department of Computer Engineering, Shahreza Campus, University of Isfahan, Iran. Electronic address: `m.lotfi@shr.ui.ac.ir`

Patient similarity assessment, which identifies patients similar to a given patient, can help improve medical care. The assessment can be performed using Electronic Medical Records (EMRs). Patient similarity measurement requires converting heterogeneous EMRs into comparable formats to calculate their distance. While versatile document representation learning methods have been developed in recent years, it is still unclear how complex EMR data should be processed to create the most useful patient representations. This study presents a new data representation method for EMRs that takes the information in clinical narratives into account. To address the limitations of previous approaches in handling complex parts of EMR data, an unsupervised method is proposed for building a patient representation, which integrates unstructured data with structured data extracted from patients' EMRs. In order to model the extracted data, we employed a tree structure that captures the temporal relations of multiple medical events from EMR. We processed clinical notes to extract symptoms, signs, and diseases using different tools such as medspaCy, MetaMap, and scispaCy and mapped entities to the Unified Medical Language System (UMLS). After creating a tree data structure, we utilized two novel relabeling methods for the non-leaf nodes of the tree to capture two temporal aspects of the extracted events. By traversing the tree, we generated a sequence that could create an embedding vector for each patient. The comprehensive evaluation of the proposed method for patient similarity and mortality prediction tasks demonstrated that our proposed model leads to lower mean squared error (MSE), higher precision, and normalized discounted cumulative gain (NDCG) relative to baselines.

# 1 Introduction

The patient similarity assessment identifies patients similar to a given patient. It allows physicians to gain insights into the records of matching patients and provide better treatments. Calculating patient similarity requires measuring the distance between patients within a population (1). A distance could be calculated based on various structured and unstructured data types in an electronic medical record (EMR).

EMRs can be processed in the same way as general documents modeled as sequences of words. The difference is that EMRs are sequences of patient events, such as diagnoses, procedures, and medications. The representation of an EMR is a low-dimension and fixed-length embedding vector, so it can be used as an indicator to measure similarity between patients, simply like a representation of a document that can be applied to measure similarity between notes.

Among previous works on patient representations based on EMRs, some have relied on structured data types (2–6), while others have only used unstructured data (7,8). In some research studies, structured and unstructured data have been represented in separate vectors, and different parts of the EMRs have not been integrated before representation (9,10).

Registration of structured data types is based on a controlled vocabulary. On the other hand, unstructured data is written in natural languages. Medications and diagnoses are an example of structured data, while clinical notes like discharge sheets are unstructured data in EMRs. Many valuable patient details, such as past illnesses, signs, and symptoms, are recorded only in clinical notes.

This paper presents a new patient representation model called UTTree (Unstructured Temporal Tree), which integrates unstructured and structured parts of EMRs. This model benefits from the Unified Medical Language System (UMLS) for extracting biomedical entities from clinical notes. We used the tree structure to model different parts of EMR. The tree structure helps to show temporal aspects, such as the co-occurrence of medical entities. We also developed an extended version of the original model, called UTTree-H (Unstructured Temporal Tree with History), to emphasize the previously detected diseases in the past medical history of patients that could influence their current health conditions. The presence of current and previous medical events was controlled using the relabeling approaches in tree construction. The traversing of the tree led to the sequences maintaining the temporal relations of medical events. These sequences were used as inputs to the representation algorithm, leading to embedding vectors, which can be used in downstream processes, such as the patient similarity assessment. The proposed model was evaluated by calculating the reduced mean squared error (MSE), precision, and normalized discounted cumulative gain (NDCG). According to the results, our model outperformed other baselines.

The contributions of the study are summarized in the following:

- UTTree: Proposing a new patient representation based on integrating unstructured and structured data in EMR modeled as a tree-based structure; applying new relabeling approaches on the tree to enrich the generated sequences through traversing the tree

- UTTree-H: Enhancing the model by considering a broader temporal aspect, including the medical events extracted from the patients' past medical history in EMR
- Evaluating the produced embedding vectors on downstream tasks, including patient similarity assessment and mortality prediction

The remainder of the paper is organized as follows. The second section summarizes relevant works and reviews the history of patient representation. Our patient representation method is described in the third section. The fourth section discusses the results. Eventually, the conclusion and potential future research are summarized in the fifth section.

## 2  Related work

In this section, we provide a classification of patient representation methods from different perspectives, such as data types, information aggregation strategies, time modeling methods, similarity metrics, downstream task types, and technologies:

Data Type: This category is based on the type of data used for generating patient representations. Some studies used only structured data in EMR, such as diagnostic codes or drugs and laboratories, recorded using terminology systems. Choi et al. (4) used the Word2Vec method (11,12) to represent diagnostic codes. Furthermore, Choi et al. (13) combined the vector resulting from the representation of the diagnostic code with the representation of the patient. This model was used to predict heart failure. The representations created by Miotto et al. (10) were based on clinical notes and structured data. Their results outperformed those of Choi et al. (13) in disease prediction. Miotto et al. (10) used the autoencoder method for representation. However, Choi et al. (13) showed that the autoencoder method is not better than other baselines in predicting diseases when only structured data is used. Using these results, we can infer the importance of clinical notes in creating a patient representation of EMRs.

Alternatively, some studies utilize clinical notes to represent patients. Dligach et al. (14) reported that clinical notes could produce results comparable to those of other methods. The time-aware patient EHR representation (TAPER) proposed by Darabi et al.(9) represents clinical notes using a transformer model. It was shown that clinical notes combined with structured data outperformed those used alone. These conflicting results suggest the need for further studies on the benefits of combining structured data and unstructured clinical notes.

Previous studies used different approaches to preprocessing, which can be divided into two groups regarding clinical notes. The first group involved named entity recognition (NER), annotation, and normalization using UMLS (14, 15), while the second group did not use normalization (7, 9). The NER identifies and aligns biomedical entities to knowledge bases, enhancing decision-making efficiency (16). Previous research has not verified the effects of NER and normalization on the patient similarity assessment using UMLS.

A variety of text representation methods, such as Doc2Vec (6), topic modeling (10), autoencoder (7), and pre-trained BERT (9), have been used in studies. The TAPER method (9), which uses transformer-based models on clinical notes, has shown better results than topic modeling (10).

However, given the variety of datasets used, the way medical codes are grouped, and the diversity of the tasks and ablation evaluations, it is difficult to compare their results.

Fusion strategy: Data fusion of multiple sources yields more accurate, more consistent, and more helpful information than any single data source can provide. Data fusion processes are often categorized as late and early (17). The kind of fusion strategy is another issue that has not been discussed in many previous studies (18). The majority of previous studies have employed late fusion, during which information from different types of data is represented separately. For example, in (9), a separate representation vector was produced for each datum in the EMR. Then, the resulting embedding vectors were concatenated to form the ultimate patient representation. However, in early fusion, data is integrated before being represented. Henriksson et al. (15) showed that in adverse drug reaction (ADR) identification, an early fusion strategy performs better than late fusion.

Time modeling methods: Regardless of the type of data and fusion strategy, modeling time data is one of the essential aspects we evaluated in previous studies. Based on how they are modeled in the time dimension, studies can be divided into four categories:

1. Studies that use medical events regardless of their timing (7, 10, 14)
2. Studies that take into account the chronological order of events
3. Studies that consider time irregularities, along with chronology, by accounting for different approaches (4, 6, 19-23) (Windows intervals vary from five minutes to one year.)
4. Studies that place importance on recent events (19)

The first group cannot provide a comprehensive case explanation regarding the importance of prioritization and latency in medical decision-making. In the second group of studies, the interval between events is not taken into account. The third group uses a time window and normalizes the event duration based on its length. Given the wide range of time intervals (from five minutes to one year), a comparison is impossible. The chronic or effective diseases that have occurred in the past and can influence the patients' present conditions are not considered.

Similarity metrics: The patient representation has been applied to various applications and downstream tasks. One such task is patient similarity. This sub-section explores how patient similarity has been measured in previous research. Methods in which similarity is determined by measuring the distance between embedding vectors of different data types can be found in studies like (24-28). However, determining patient similarity according to patient representation has received less attention. Pokharel et al. (6) used a combination of structured data in a tree model. They presented an evaluation method based on final diagnoses. This method is inspired by Gottlieb et al. (20). However, it considers the priority of the final diagnosis recorded for a patient. We have followed the same approach in this study. The most used similarity metric was the calculation of cosine (6, 10), Euclidean (21), and Mahalanobis (22) distances. Comparing the performance of different metrics in the patient similarity assessment requires further investigation.

A review of the methods reveals that representing patient information has received attention from researchers. Aside from the aspects mentioned above, the studies differ in the downstream tasks and the technology used. However, different datasets and experimental designs make it difficult to compare the reported results. Table 1 presents some characteristics of studies in this area.

Table 1. A comparison of the characteristics of the studies on the patient representation

| Reference Number | year | Supervised(S), Learning manner | Structured data | | | | | | | Unstructured data | Method | Downstream Tasks (Evaluation metric, result) | Considerations | | |
|---|---|---|---|---|---|---|---|---|---|---|---|---|---|---|---|
| | | | Demographics | Vital Signs | Diagnosis | Medication | Laboratory | Procedures | | | | | Time order based | UMLS alignment | Fusion strategy |
| (19) | 2015 | S | ✗ | ✗ | ✓ | ✗ | ✗ | ✗ | ✗ | Deep Boltzmann machine | Prediction: (Disease, only utilizes 70 most frequent ICD-9 codes) | ✓ | ✗ | L |
| (10), Deep Patient | 2016 | U | ✗ | ✗ | ✓ | ✓ | ✓ | ✓ | ✓ | Autoencoder | Prediction: (Disease, AUC=0.773) | ✗ | ✗ | L |
| (13), Med2vec | 2016 | S | | | ✓ | ✓ | ✗ | ✓ | ✗ | Word2Vec | AUC, Recall@30 | ✗ | ✗ | L |
| (20) | 2016 | U | ✗ | ✗ | ✓ | ✓ | ✓ | ✗ | ✗ | Word2Vec | Prediction: (Disease, on 80 diagnosis AUC =0.67) | ✓ | ✗ | E |
| (31) | 2016 | U | ✗ | ✗ | ✓ | ✓ | ✓ | ✗ | ✗ | Word2Vec | Prediction: (Disease -heart failure, AUC =0.82) | ✓ | ✗ | L |
| (15) | 2016 | U | ✗ | ✗ | ✓ | ✓ | ✗ | ✗ | ✗ | Word2Vec | Prediction: (ADE-related diagnoses, AUC=0.93, Accuracy = 0.84) | ✓ | ✗ | L |
| (23) | 2017 | U | ✗ | ✗ | ✓ | ✓ | ✓ | ✗ | ✗ | CNN Based | Prediction: (Readmission, AUC = 0.750) | ✓ | ✗ | L |
| (32) | 2017 | U | ✗ | ✗ | ✓ | ✓ | ✓ | ✗ | ✗ | RNN-Based | Prediction: (Disease (heart failure), AUC =0.883) | ✓ | ✗ | L |
| (14) | 2018 | S | ✗ | ✗ | ✗ | ✗ | ✗ | ✗ | ✓ | Autoencoder | Prediction: (Billing code, Precision=0.709, Recall=0.725, F1=0.715) | ✗ | ✓ | L |
| (4), Patient2Vec | 2018 | S | ✓ | ✗ | ✓ | ✓ | ✓ | ✗ | ✗ | RNN-Based | Prediction: (Hospitalization, AUC = 0.799) | ✓ | ✗ | L |
| (7) | 2018 | U | ✗ | ✗ | ✗ | ✗ | ✗ | ✗ | ✓ | Autoencoder + Doc2vec model. | Prediction: (Mortality In hos AUC=0.93), (30 days AUC=0.81) | ✗ | ✗ | L |
| (33) | 2018 | U | ✗ | ✗ | ✓ | ✓ | ✓ | ✗ | ✗ | Document-level embedding. | Disease (breast cancer) prediction:(AUC=0.83, MSE=0.005) | ✓ | ✗ | L |

| | | | | | | | | | | | | No use of preprocessing. | | | |
|---|---|---|---|---|---|---|---|---|---|---|---|---|---|---|---|
| (9) TAPER | 2020 | U | ✗ | ✗ | ✓ | ✓ | ✓ | ✓ | ✓ | | | Transformer-based | Prediction: (Mortality AUC-ROC=0.63, PR-AUC=0.65) Prediction: (Hospitalization, AUC-ROC=0.67, PR-AUC=0.68) | ✓ ✗ | L |
| (34), EHR2vec | 2020 | U | ✗ | ✗ | ✗ | ✗ | ✗ | ✗ | ✓ | | | self-attention mechanism | Prediction: (Disease_SLE) | ✓ ✗ | L |
| (21) | 2020 | U | ✓ | ✓ | ✓ | ✓ | ✓ | ✓ | ✗ | | | RNN-Based | Prediction: (Disease _sepsis, AUROC= 0.786, MAP= 0.797) | ✓ ✗ | E |
| (6),Temporal Tree | 2020 | U | ✗ | ✗ | ✓ | ✓ | ✓ | ✓ | ✗ | | | Doc2Vec | Patient Similarity: (MSE@1 = 0.267, DCG@1=0.477, Precision@1 = 0.698) | ✓ ✗ | E |

A review of the previous studies indicates that although they have addressed different aspects of patient representation, there are still issues such as the following that have not been studied:

- In most previous studies, structured and unstructured information representations were created in two separate vectors and then combined. For the first time, this study mixes the structured and unstructured data through the tree structure to produce one representation vector for both.
- Most previous studies employed general methods of text preprocessing for clinical notes. Fewer studies used external knowledge sources, such as UMLS, in the clinical processing of notes.
- Previous studies have given the same weight to all parts of clinical notes and have not explored the significance of different parts of the text in producing EMRs. However, the textual information in a patient's electronic medical record contains information about their family history, illness history, and current referral, which do not have the same importance.

These challenges prompted us to develop a model for future research without the mentioned problems.

## 3  Methodology

This study proposes a model called UTTree (Unstructured Temporal Tree) to represent patient information. The model combines the structured data and textual information of the EMR recorded in the latest visit of the patient using a tree structure, as described in Section 3.1. Additionally, the model is enhanced by the patients' medical history in their EMR. Section 3.2 explains the extended model, UTTree-H (Unstructured Temporal Tree with History). The two methods are different in the way they implement the relabeling mechanism. This difference can be found in the workflow shown in Fig. 1 (steps 2-b and 2-c).

### 3.1  The UTTree

A goal of this model is to combine structured and unstructured data and give the combined data as an input to the document representation algorithm. We combined the data using a tree structure. The following gives an explanation of how trees are created.

The input of the model is a database containing the various types of information recorded in the EMRs of several patients, and its output is the degree of similarity between every two patients. According to the proposed algorithm, the degree of similarity is calculated by comparing the cosine distance of the embedded vectors generated for each patient. The output of the proposed algorithm is compared with the ideal similarity scores. This section continues with an explanation of the different parts of the workflow. Fig. 1 depicts the main steps of the algorithm in a workflow diagram.

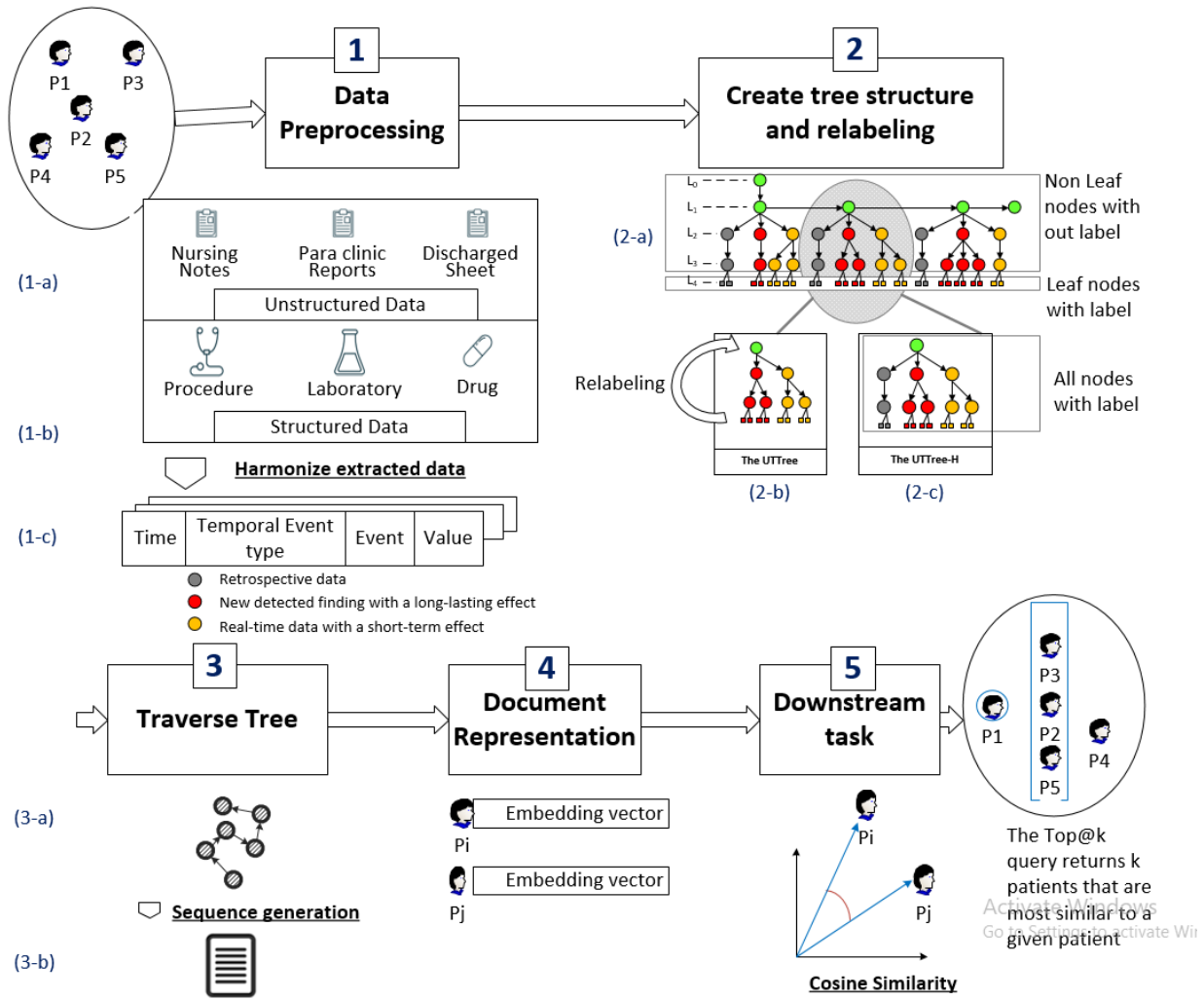

Fig. 1. Overview of the proposed method. First of all, in step 1, EMR data for each patient is taken from the dataset (1-a). Afterward, structured and unstructured data are preprocessed separately (1-b). The preprocessed data are homogenized using a data structure called quadruple (1-c). In step 2, quadruples are used to develop a tree structure (2-a). Next, two kinds of relabeling processes are conducted to assign labels to non-leaf nodes. The UTTree model uses only new detected findings (red nodes) and real-time data (yellow nodes) (2-b) for relabeling the non-leaf nodes in the middle level of the tree, while the UTTree-H model uses the retrospective data, as well, for assigning labels to non-leaf nodes. The output of the second phase is a tree structure for each patient, in which all nodes have labels. In step 3, traversing the tree (3-a) provides rich sequences (3-b), which are used as inputs for step 4. In step 4 the sequences are represented through the document representation algorithm to generate embedding vectors for patients. Finally, the cosine distance between the vectors is considered the similarity score. For each patient, the most similar K patients are returned (5).

Data preprocessing: In the first phase, the goal is to identify a person's previous and current diseases from the descriptions written in clinical texts. Since signs and symptoms describe the diseases in medical texts, all terms related to diseases, signs, and symptoms are extracted and grouped according to their section headers in medical notes. After section detection, the algorithm employs UMLS to extract concepts for each expression. The information about prescriptions and laboratory tests is also extracted. A quadruple data structure is used to homogenize the structure of information extracted from structured and unstructured sources that provide the basis for combining information.

Creating the tree structure and relabeling: The quadruples are applied in developing a temporal tree model. We create a subtree for each time window. There are branches for each data type (disease, prescription, laboratory tests). In each branch of each subtree, the medical events related to that time window and datatype are used as leaf nodes. In the tree's construction, a modified model of the Weisfeiler-Lehman graph kernels (23) relabeling method is used to compare graphs that capture topological and label information. Using these kernel co-occurrences, the relation between levees in each subtree (medical events) converts to rich compounds in non-leaf nodes. In this step, two relabeling methods are used. In the first method, the extracted medical events related to the same visit are used in relabeling non-leaf nodes. In the second method described in section UTTree-H, data related to the patients' medical history are also used.

Traversing the tree: In the third step, traversing the temporal tree using the BFS method yields sequences, which are used as inputs of the embedding algorithm to represent patient information as vectors.

Document representation: The result of tree traversal is a sequence of components, i.e., a new document for each patient. This document serves as the input of the representation algorithm. This study employs the Doc2Vec algorithm (Le and Mikolov, 2014), which is a sentence embedding method (24) used to create a vectorized representation of a group of words taken collectively as a single unit, rather than giving only the simple average of the words in the sentence.

Downstream task: The embedding vectors are fed to the patient similarity assessment model. It works based on the cosine distance between patient embedding vectors and the gold standard to measure model accuracy. The results obtained from this method are compared with several baselines. The details of each stage are presented in the following.

### 3.1.1 Data preprocessing

Unstructured data processing: We define the process of automatically extracting clinical concepts from unstructured clinical notes in EMRs. A large amount of the information in an electronic patient record system is unstructured in the form of free text (25). Clinical text is written by various professionals such as physicians, nurses, physiotherapists, and psychologists. It is often written under time pressure and contains misspellings, non-standard abbreviations, jargon, and incomplete sentences. Therefore, it is difficult to employ natural language processing (NLP) tools mainly developed for other text types (26).

Clinical notes in EMRs contain different text types, such as discharge notes, nursing reports, and pharmacy notes. Each of the standard clinical notes, especially discharges notes, consists of several components or sections (27). However, there is significant variation in sections and the descriptive phrases in section headers. This challenge in the NLP of clinical notes is named flexible formatting (37, 40). In this study, we consider the effect of the patients' medical history in interpreting their new clinical conditions. Therefore, it is essential to differentiate between the patients' medical history (previous active or resolved medical conditions) and new findings in the first step. After sectionizing each note by medspaCy (28), we follow the next step, which is named-entity recognition (NER). In this step, medical named entities, such as diseases, are recognized from medical texts. This step is necessary for representing a complete overview of a patient's medical history and current significant findings. We use MetaMap (29) for this step.

Finally, we perform concept encoding to map the extracted terms to standard terminologies. A doctor may express the same disease differently, so mapping based on ontology is crucial. In this step, different forms of one disease are automatically mapped to one preferred name. In this process, associated concepts are retrieved from MetaMap (42) and sorted according to their similarity to the searched term. Each term is assigned the concept with the highest score returned by MetaMap. After the concept mapping, the extracted concepts are sorted by registration time. During step (1-a), the input notes are batched into a set of records. Each record includes the original phrase, note type (discharge, nursing, reporting.), section category, registration time, negation tag, semantic type, and concept ID of each extracted phrase. The tools and libraries used in this process are listed in Appendix A.1.

Structured data processing: EMRs contain several structured data types. This study uses data on prescriptions, laboratories, and procedures. Structured data types are captured using controlled vocabularies rather than narrative text. Healthcare providers usually use the LOINC[1] standard for identifying laboratory tests or the RxNorm[2] standard for identifying clinical drugs. Lab test results are usually documented, as well. Depending on age, gender, and some other characteristics, this result is interpreted as low, usual, or high. In this study, we use these categories rather than the original values.

Harmonization of extracted data: To harmonize extracted information from clinical notes and structured parts of EMR, we introduce a flat format called quadruple. The quadruple data structure format consists of four components listed in Table. 2: time (t), event type (y), event (e), and value (v), which are represented by $\{t_i, y_i, e_i, v_i\}$, $1 \leq i \leq n$. The parameter $n$ is the total number of extracted information. Patients can have different numbers of quadruples.

Table 2. Components of quadruple

| Component | Definition |
| --- | --- |
| Time - $t_i$ | The timestamp that each clinical event registers in the EMR. |
| Temporal Event role - $y_i$ | The following temporal roles were taken into consideration in extracted disease in EMR: |

---

[1] Logical Observation Identifiers Names and Codes

[2] https://www.nlm.nih.gov/research/umls/rxnorm/index.html

| | |
|---|---|
| | Retrospective data (a): This role is assigned to the concepts extracted by processing medical history texts. |
| | New detected finding with a long-lasting effect (b): every disease that registers in current visit. We hypothesized that disease has more lasting effects than other medical events. |
| | Real-time data with a short-term effect (c): The parameter used to model medical events other than a disease, such as laboratory results and prescribed drugs. These items are updated more frequently than disease events. Thus, they are only used to create subtrees (time windows) in which they are recorded. |
| Event - $e_i$ | The extractable medical event type (disease, sign, symptom, drug, lab, procedure.) |
| Value - $v_i$ | The extractable medical event value |

### 3.1.2 Creation of tree structure and relabeling

This phase includes two steps: creating a tree and relabeling non-leaf nodes by Weisfeiler-Lehman graph kernels (23).

Creating the tree: A four-level tree is created for each patient rooted by the patient identifier.

Each tree has subtrees at its second level. The number of subtrees is equal to the number of nonempty time windows. The user defines the length of the time window. Herein, we use a one-day time window. According to Table 2, the number of included quadruples in each time window is determined by the time field. The algorithm creates branches in subtrees for each nonempty temporal event role at the third level of the tree. As mentioned in Table 2, there are three temporal event roles. Thus, the third level of each subtree can have up to three different nodes. The first one is for retrospective data, the second one is for new detected findings with a long-lasting effect, and the last one is related to real-time data with a short-term impact. The fourth level adds each pair of extracted values and event types. As mentioned in Table 2, event types could be disease, sign, symptom, drug, laboratory, and procedure. By the end of this phase, we have a four-level tree in which only leaf nodes have labels. The non-leaf nodes are relabeled in the next step. The pseudo-code can be found in Appendix A.5 in the supplementary files.

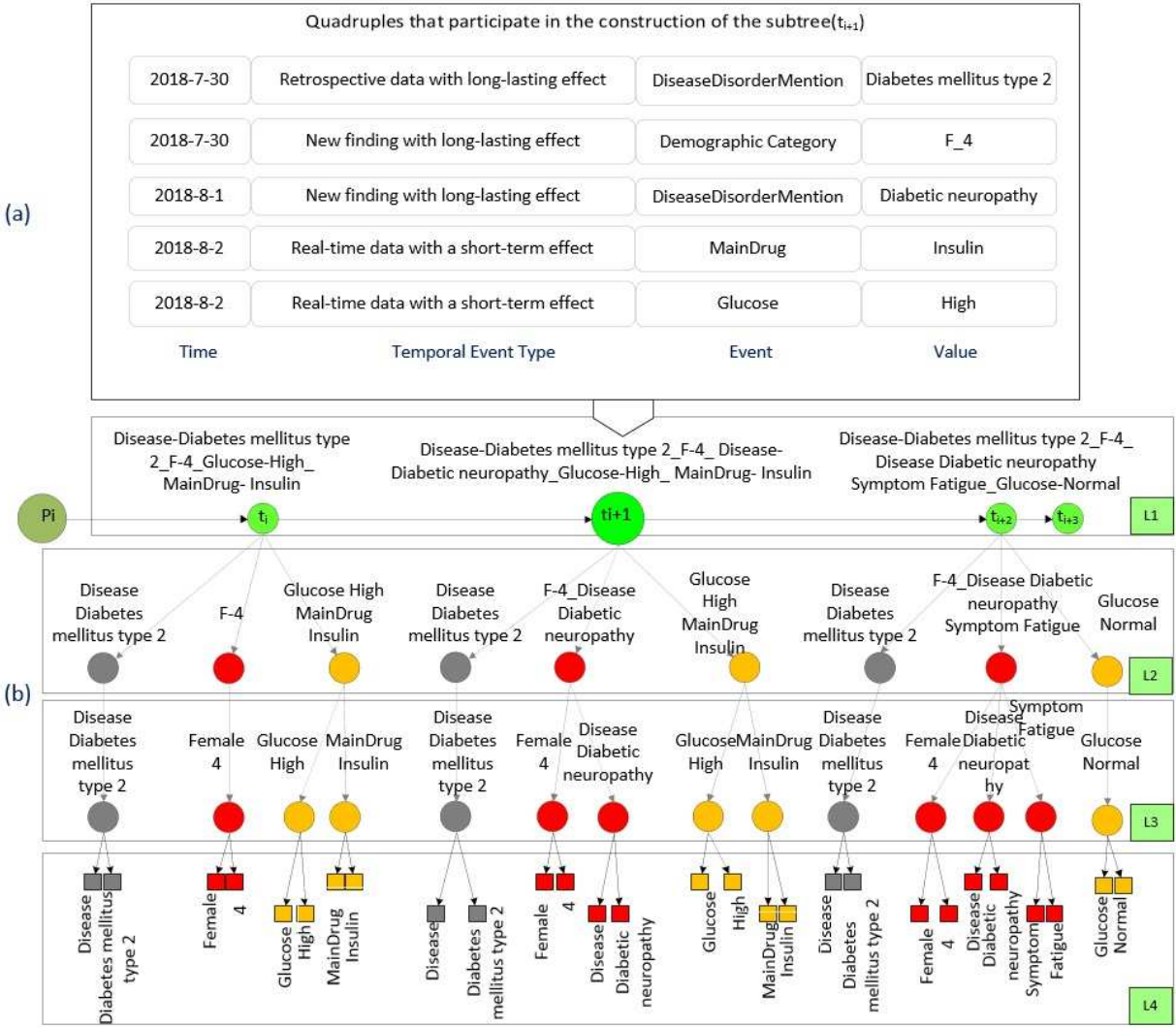

Fig. 2. An example of creating and relabeling a tree. At the top of the figure (a), we see a set of quadruples used to construct the subtree(ti+1), which contains five records. The main steps in Fig. 1 indicate that only leaf nodes on level 4 are labeled after the tree is created. Through the relabeling step, the non-leaf nodes of the higher levels are labeled in order. The UTTree method takes into account the current visit information for relabeling, while the UTTree-H method uses the past history of the patient, as well.

Relabeling: Relabeling non-leaf nodes in the tree allows us to capture the relationship between different medical events. Fig. 2 shows an example. This phase is completed using the Weisfeiler-Lehman kernel, which is designed for graphs and consists of four steps: multiset-label determination, sorting, label comparison, and relabeling. Only the leaf nodes have labels in the tree created in the previous step, and they can only be accessed through their parent nodes. As a result of two rounds of relabeling, the tree's root turns into a combination of medical events that occur within a specified time window. Fig. 2 shows the mechanisms of relabeling non-leaf nodes, focusing on Part 2-b of Fig. 1. Each subtree corresponds to a time window determined by the field "Time" in the quadruple. In the relabeling step, three approaches are used to change the labels of non-leaf nodes based on leaf nodes' labels. In the following sections, the three methods, namely the temporal tree (6), UTTree (the proposed model), and UTTree-H (the enhanced proposed model described in the next section), are applied. The temporal tree approach only relabels real-time nodes (yellow ones). In the UTTree approach, the nodes of newly detected findings (red ones) and real-time data are also used.

### 3.1.3 Traverse Tree

In this phase, we have a tree in which all leaf and non-leaf nodes have labels.

The tree can be traversed to generate sequences of terms. The traversal of a tree data structure is the process of visiting each node in the tree exactly once. It is classified by the order based on which the nodes are visited. This study uses the breadth-first search (BFS) traverse method. In breadth-first search (BFS) or level-order search, the search tree is broadened as much as possible before going to the next depth. Fig. 3 shows sequences created after tree traversing.

**Seq1**: Disease, Diabetesmellitustype2, Female,4, Glucose, High, MainDrug, Insulin, Disease, Diabetesmellitustype2, Female,4, Disease, Diabetneuropathy, Gloucose, High, MainDrug, Insulin, Disease, Diabetesmellitustype2, Female,4, Disease, Diabetneuropathy, Symptom, Fatique, Glocose, Normal

**Seq2**: DiseaseDiabetesmellitustype2, Female4, GlucoseHigh, MainDrugInsulin, DiseaseDiabetesmellitustype2, Female4, DiseaseDiabetneuropathy, GloucoseHigh, MainDrugInsulin, DiseaseDiabetesmellitustype2, Female4, DiseaseDiabetneuropathy, SymptomFatique, GlocoseNormal,

**Seq3**: DiseaseDiabetesmellitustype2, Female4, GlucoseHighMainDrugInsulin, DiseaseDiabetesmellitustype2, Female4DiseaseDiabetneuropathy, GloucoseHighMainDrugInsulin, DiseaseDiabetesmellitustype2, Female4DiseaseDiabetneuropathySymptomFatique, GlocoseNormal,

**Seq4**: DiseaseDiabetesmellitustype2Female4GlucoseHighMainDrugInsulin, DiseaseDiabetesmellitustype2Female4DiseaseDiabetneuropathyGloucoseHighMainDrugInsulin, DiseaseDiabetesmellitustype2Female4DiseaseDiabetneuropathySymptomFatiqueGlocoseNormal,

Fig. 3. Temporal sequences in the BFS order generated for the tree with four levels shown in Fig 2; During the relabeling process, all leaf and non-leaf nodes of the tree are labeled in step 3 of the workflow. A sequence is generated for each tree level when the tree is traversed in the BFS order. The sequences combine to form the final document for each patient.

### 3.1.4 Patient representation

We have now reached a stage where we can generate embedding vectors for individual patients using document representation methods. Sequences created in the previous step are represented using the document representation technique. In document representation, the semantics of a document are encapsulated in real-valued vectors, which can be manipulated in downstream tasks. Relabeling results in documents that demonstrate the co-occurrence of medical events. Doc2Vec is one of the most popular ways of representing documents.

The Doc2Vec is based on Word2Vec and uses an unsupervised learning approach to learn the document representation. The input text (in our case, medical events) per document (in our case, temporal tree) can be varied, while the output is fixed-length vectors. The Doc2Vec includes two methods. The Distributed Memory Model of Paragraph Vectors (PV-DM)(30) is similar to the continuous bag-of-words approach in Word2Vec. The other method used in Doc2Vec is the Distributed Bag-of-Words version of Paragraph Vector (PV-DBOW) (31), which is similar to the Skip-gram approach in Word2Vec. The PV-DM model computes the probability of a target word in a lexical context based on the surrounding words. The document is mathematically expressed as below in Equation 1:

$$\sum_{t \epsilon T_w} \log p(w_t | C_t, d_t) \tag{1}$$

where $w_t$ is the target word, $T_w$ is the set of training words, $C_t = [w_{t-L}, \cdots, w_{t-1}, w_{t+1}, \cdots, w_{t+L}]$ are context words that occur within a window size of L words around $w_t$, and $d_t$ denotes the document corresponding to the $t_{th}$ training instance. There are two optimization methods for Doc2Vec: hierarchical softmax and negative sampling. In this study, we use the former. The experiments providing better results for this optimization method are listed in Appendix A.3.

### 3.1.5 Downstream Tasks

Using the representations produced in this step, we perform two types of downstream tasks, i.e., calculating patient similarity and predicting patient mortality. Section 4 examines the generated representations based on this downstream task. In this study, the main objective is to evaluate patient similarity. In this task, the score of similarity with other patients is calculated and sorted in descending order for each query patient in the dataset. The top K patients on the sorted list are returned as output to the algorithm. Moreover, the generated vectors are used to predict mortality, which is a supervised classification task.

## 3.2 UTTree-H

The UTTree is modeled using data from current admissions. As a part of the relabeling process, the model UTTree-H incorporates the patients' history. Due to this difference, the final generated sequences can include a combination of current clinical events and patient history. Therefore, the final representation is influenced by the patient history. In the UTTree-H, all three types of temporal event data are used to label non-leaf nodes.

An example is provided in Fig. 4 to illustrate the importance of using concepts extracted from the past history of patients recorded in clinical notes to construct labels of nodes. We have two patients with similar information obtained from their current visits. The green boxes indicate the current data for the two patients. They have different past histories. One is pregnant, while the other has thalassemia. We have extracted information about the patients' pregnancy and thalassemia history from their clinical notes. In the UTTree-H model, patient history is incorporated into the creation of labels, resulting in a unique final sequence for each patient. Blue boxes represent the information used to construct each patient's final sequence. The detailed information (past medical history) makes a difference between the two sets.

This distinction is critical in assessing patients' similarities. Unlike iron deficiency anemia, identifying patients with thalassemia is vital for avoiding unnecessary iron replacement therapy. Iron overload can lead to several complications in patients with thalassemia (32), impair their immune system, and place them at risk of infection and illness(33).

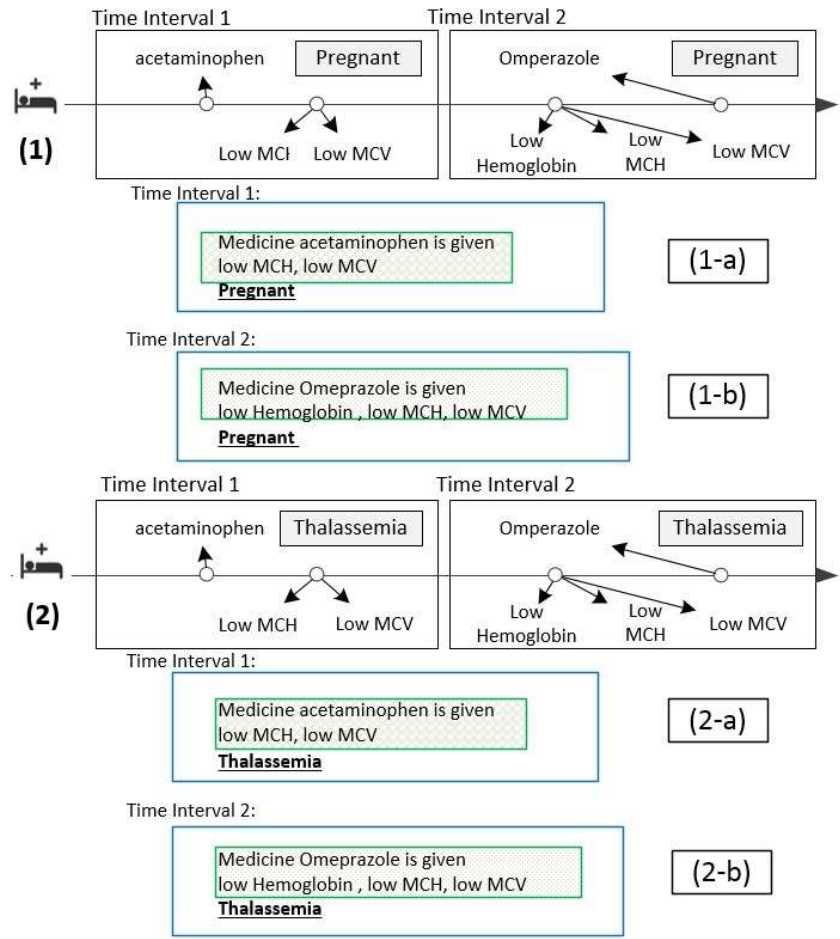

Fig. 4. The effect of paying attention to the concepts extracted from the past history section of clinical notes

## 4  Results and Evaluation

Two factors impact the quality of our proposed patient representation method: the performance of the text representation algorithm and the relabeling method used (please see Appendix A.3 in the supplementary file for the hyperparameter selection). Since both factors contribute to the construction of patient representations, the effects of both of them are examined in the following experiments. Section 4.1 describes the experimental setting, and section 4.2 reviews the utilized dataset. Section 4.3 analyzes the effect of integrating unstructured and structured data in EMR in generating patient representation through the UTTree model. The resulting embedding vectors generated by the UTTree model are examined in the patient similarity task.

Section 4.4 examines the effect of additional temporal aspects, including the medical events extracted from the patients' past medical history in the UTTree-H model. In this section, we study under what conditions the use of past medical history records in constructing the UTTree-H could improve performance. The generated vectors are evaluated in a patient similarity task. In the last experiment, reported in section 4.5, we seek to find out whether the generated embedding vectors are also effective in problems other than patient similarity.

## 4.1 Experimental setting

This section explains the approaches used for creating the gold standard. In the following, we explain the evaluation criteria. The effectiveness of the proposed representation method is evaluated in two types of downstream tasks: (i) similar patients' retrieval and (ii) prediction model. Each task has its specific evaluation criteria.

### 4.1.1 Gold standard

We use all available final diagnosis codes for evaluation and assign a weight to each based on diagnosis code priority in each patient's EMR. To calculate the similarity between patients *A* and *B*, we use Equation 2 (SimIndex), defined by Pokharel et al. (34):

$$SimIndex(A, B) = \frac{\sum_{i=1}^{N} min(a_{w_i(p)}, b_{w_i(p)})}{avg\left(\sum_{i=1}^{N} a_{w_i(p)}, \sum_{i=1}^{N} b_{w_i(p)}\right)} \quad (2)$$

In the equation above, $a_{w_i(p)}$ and $b_{w_i(p)}$ are the weights of diagnosis *i* with priority *p* expressed as an ICD9 disease code for patients *A* and *B*, respectively.

Fig. 5 illustrates the evaluation task. The similarity between two patients can be determined by the number of diagnoses shared by the two patients and the corresponding rankings of the common diagnoses. The diagnoses at higher ranks are given greater weights than those at lower ranks. We use three evaluation metrics in this study.

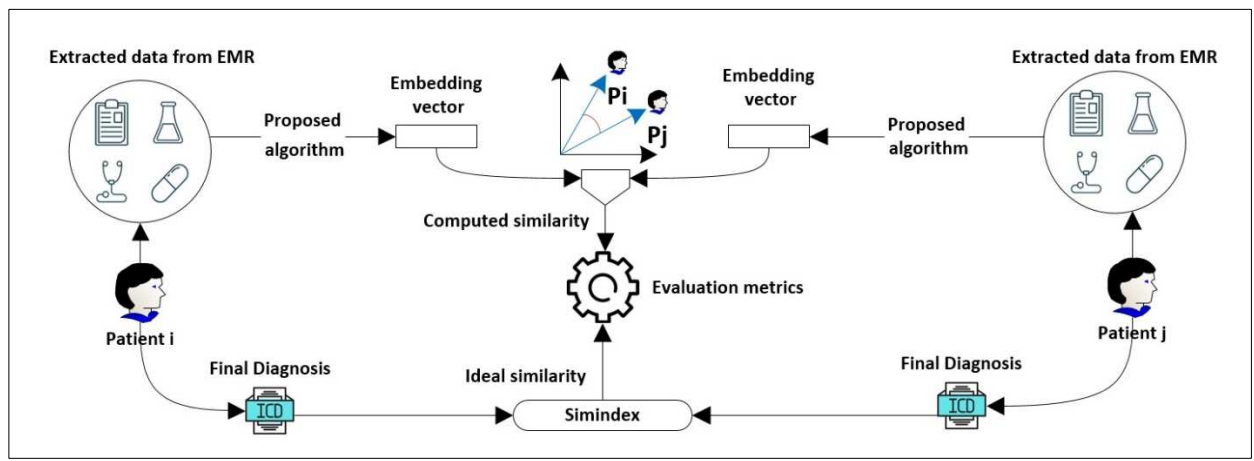

Fig. 5. Evaluation of the similarity calculated based on the gold standard

### 4.1.2 Evaluation metrics in patient similarity

We evaluate the proposed method in the retrieval of similar patients using the following evaluation metrics:

Mean Square Error (MSE@K): The prediction error is the difference between the actual and predicted values (35). The Top@k query returns K patients most similar to a given patient, based on the cosine similarity measure of the embedding vectors.

$$\left(\frac{1}{n}\right) \sum (actual - predicted)^2 \qquad (3)$$

In this equation, *n* is the number of patients in the dataset, and Σ is the summation notation. Actual values are determined by observing the K most relevant patients in the gold standard for each patient. The highest similarity predicted for each patient is compared with their similarity in the gold standard.

Normalized Discounted Cumulative Gain (nDCG@K):

This measure is used for determining the ranking quality. According to this idea, products that a user likes should be ranked first on the recommendation list. DCG (Discounted Cumulative Gain) normalization is computed as:

$$nDCG_p = \frac{DCG_p}{IDCG_p} \qquad (4)$$

Where IDCG is ideal discounted cumulative gain,

$$IDCG_p = \sum_i^{REL_p} \frac{rel_i}{\log_2(i+1)} \qquad (5)$$

where $REL_p$ is a sorted list of relevant documents, i is the current result's position number, and IDCG is the largest DCG in the ideal state. The DCG measures the best-ranking result, and nDCG is the DCG normalized by the ideal DCG. Thus, the nDCG measure is always a number within [0, 1] (36). The nDCG is computed for the Top@k patient rankings retrieved from a query patient. The optimal Top@k ranking for a query patient is calculated according to the gold standard. We use an ordinary function for calculating the discount. For a query patient, the calculated SimIndex scores for both the ideal and actual rankings are discounted by a function of the rank position. More information about the SimIndex scores and the gold standard is provided in Section 4.1.1.

Precision@K: Using (20) as a guide, we calculated the precision of our predictions by counting the number of patients with the same top predicted discharge code as a correct diagnosis. By setting a rank threshold of K, the patient is considered relevant if any of the top K query diagnoses appear in a retrieved patient list.

## 4.2   Dataset

We use a subset of The Medical Information Mart for Intensive Care III (MIMIC-III) dataset (37). MIMIC III is a large publicly-available clinical database containing data from approximately 40,000 de-identified patients. These data come from patients admitted to a medical center in Boston, Massachusetts, from 2001 to 2012. Each record in the dataset includes ICD-9 codes, identifying diagnoses, and performed procedures. The dataset consists of 112,000 clinical report records (average length: 709.3 tokens). We select a subset of patients whose final diagnosis is one of the top ten most frequent first diagnoses in the whole dataset. Records with no laboratory tests or prescriptions are excluded. This collection includes over 7300 admissions. The average number of words in medical notes is about 7800, and the average number of visits per patient is 1.29.

### 4.3 Evaluation of UTTree

In the following evaluation, we seek to answer this question: Can the integration of structured and unstructured data in an EMR that generates patient representations through the UTTree model lead to effective results? The UTTree relabeling model is compared with the temporal tree model (6). Together with the structured data, the UTTree model uses the disease concepts extracted from the documents. The BFS is used for traversing the tree that leads to sequences. The Doc2Vec algorithm represents the resulting sequences, and then the embedding vectors are used in the downstream similarity calculation. The distance of embedding vectors is calculated using a cosine metric and compared with the gold standard as described in section 4.1.1 by means of MSE, nDCG, and precision for retrieving 1, 5, 10, and 20 similar patients. Table. 3 lists the results. The proposed model, UTTree, is implemented using two Doc2Vec methods, namely PV-DBOW and PV-DM (number one and number two), and is compared to the following baselines (numbers four to six):

1. UTTree_[PV-DBOW]: This implementation employs UTTree relabeling, and after traversing the tree, the resulting sequences are represented by the Doc2Vec (Distributed Bag-Of-Words (DBOW) Model, similar to the skip-gram model of Word2Vec, which guesses the context words from a target word).
2. UTTree_[PV-DM]: This implementation uses UTTree relabeling, resulting in sequences represented by Doc2Vec PV-DM.
3. TFIDF_Structured data: The TFIDF (term frequency-inverse document frequency) (38) is used to find non-negative vectors from the documents based on the term frequency, which indicates how frequently a term occurs in a document, and the inverse document frequency, which indicates the frequency of a term across a collection. We use only structured data in constructing the tree structure.
4. TFIDF_ Structured and Unstructured data: Same as the previous baseline, the representation is TFIDF, but we use structured and unstructured data of one visit.
5. The Temporal Tree_[PV-DBOW]: This baseline is the implementation of (6) that resulted in sequences represented by Doc2Vec (PV-DBOW).
6. The Temporal Tree_[PV-DM]: This baseline is the implementation of (6) that resulted in sequences represented by Doc2Vec (Distributed Memory (DM) that is similar to the Continuous-Bag-of-Words (CBOW) model in Word2Vec, which attempts to guess the output (target word) from its neighboring words (context words) with the addition of a paragraph ID that is the tree ID in this case).

Table 3 The effect of using clinical notes in constructing the model

| | MSE | | | |
|---|---|---|---|---|
| | MSE@1 | MSE@5 | MSE@10 | MSE@20 |

|  | MSE | | | |
| --- | --- | --- | --- | --- |
| TFIDF_Structured data | 0.320 | 0.129 | 0.080 | 0.052 |
| TFIDF_ Structured and Unstructured data | 0.310 | 0.128 | 0.078 | 0.050 |
| The Temporal Tree_ [PV-DBOW] | 0.283 | 0.098 | 0.062 | 0.044 |
| UTTree_ [PV-DBOW] | 0.239 | 0.096 | 0.069 | 0.045 |
| The Temporal Tree_ [PV-DM] | 0.280 | 0.095 | 0.058 | 0.041 |
| UTTree_ [PV-DM] | 0.235 | 0.091 | 0.064 | 0.044 |
|  | nDCG | | | |
|  | nDCG@1 | nDCG@5 | nDCG@10 | nDCG@20 |
| TFIDF_Structured data | 0.41 | 0.408 | 0.407 | 0.407 |
| TFIDF_Structured and Unstructured data | 0.421 | 0.419 | 0.412 | 0.406 |
| The Temporal Tree_[PV-DBOW] | 0.436 | 0.431 | 0.428 | 0.425 |
| UTTree _ [PV-DBOW] | 0.461 | 0.452 | 0.441 | 0.440 |
| The Temporal Tree_ [PV-DM] | 0.481 | 0.471 | 0.449 | 0.431 |
| UTTree _ [PV-DM] | 0.485 | 0.473 | 0.445 | 0.438 |
|  | Precision | | | |
|  | Precision@1 | Precision @5 | Precision @10 | Precision @20 |
| TFIDF_Structured data | 0.554 | 0.551 | 0.542 | 0.541 |
| TFIDF_ Structured and Unstructured data | 0.561 | 0.560 | 0.543 | 0.546 |
| The Temporal Tree _ [PV-DBOW] | 0.530 | 0.515 | 0.542 | 0.509 |
| UTTree _ [PV-DBOW] | 0.609 | 0.601 | 0.589 | 0.579 |
| The Temporal Tree _ [PV-DM] | 0.599 | 0.560 | 0.580 | 0.520 |
| UTTree_ [PV-DM] | 0.614 | 0.606 | 0.597 | 0.583 |

This experiment found that using unstructured data in the tree structure construction improved the accuracy of representing EMR by a tree. The following findings were obtained concerning the analysis of evaluation metrics. As for MSE, it is evident that it decreases by increasing K. The reason is that by retrieving a more extensive set of patients, there is a greater likelihood that at least one is more similar to the query patient. Another promising finding of MSE was that using unstructured data yielded a less-than-average MSE.

Furthermore, we found that the nDCG metric decreases for higher values of K. This finding confirms the previous research (6). As the set of retrieved responses grows, the likelihood of patients who do not look like the query patient increases. As a result, a more extensive collection of retrieved patients is more likely to contain unrelated values. We found that the precision decreases with the rise in K. In other words, the probability of retrieving irrelevant patients is higher with a larger K. According to the results, the implementation method PV-DM outperforms PV-DBOW. A description of the selection of hyperparameters is provided in Appendix A.3.

## 4.4 Evaluation of UTTree-H

In this section, we verify to what extent additional temporal aspects, including the medical events extracted from a patient's past medical history in EMR in the UTTree-H model, could be effective. This section aims to evaluate the performance of the enriched method UTTree-H. The UTTree-H model uses concepts extracted from the past medical history section of discharged sheets in relabeling non-leaf nodes. UTTree-H is implemented using PV-DBOW and PV-DM (numbers one and two). Evaluations are conducted using the other methods mentioned in points three to five in the following. Table 4 shows the comparison.

1. UTTree-H _ [PV-DBOW]
2. UTTree-H _ [PV-DM]
3. TFIDF with Structured and Unstructured data
4. UTTree _ [PV-DBOW]
5. UTTree _ [PV-DM]

Table 4  Evaluation of the UTTree-H relabeling method

| | MSE | | | |
|---|---|---|---|---|
| | MSE@1 | MSE@5 | MSE@10 | MSE@20 |
| TFIDF_Structured and Unstructured data | 0.310 | 0.128 | 0.078 | 0.050 |
| UTTree _ [PV-DBOW] | 0.239 | 0.096 | 0.069 | 0.045 |
| UTTree-H _ [PV-DBOW] | 0.238 | 0.094 | 0.061 | 0.039 |
| UTTree _ [PV-DM] | 0.235 | 0.091 | 0.064 | 0.040 |
| UTTree-H _ [PV-DM] | 0.232 | 0.093 | 0.061 | 0.038 |
| | nDCG | | | |
| | nDCG@1 | nDCG@5 | nDCG@10 | nDCG@20 |
| TFIDF_Structured and Unstructured data | 0.421 | 0.419 | 0.412 | 0.406 |
| UTTree _ [PV-DBOW] | 0.481 | 0.471 | 0.449 | 0.431 |
| UTTree-H _ [PV-DBOW] | 0.491 | 0.481 | 0.449 | 0.437 |
| UTTree _ [PV-DM] | 0.495 | 0.483 | 0.455 | 0.438 |
| UTTree-H _ [PV-DM] | 0.492 | 0.482 | 0.454 | 0.437 |
| | Precision | | | |
| | Precision @1 | Precision @5 | Precision @10 | Precision @20 |
| TFIDF_Structured and Unstructured data | 0.561 | 0.560 | 0.543 | 0.546 |
| UTTree _ [PV-DBOW] | 0.609 | 0.601 | 0.589 | 0.579 |
| UTTree-H _ [PV-DBOW] | 0.612 | 0.605 | 0.597 | 0.590 * |
| UTTree _ [PV-DM] | 0.614 | 0.606 | 0.597 | 0.583 |
| UTTree-H _ [PV-DM] | 0.613 | 0.607 | 0.599 | 0.592 |

The preliminary examination of the results shows that, in general, the UTTree still produces better results. The UTTree-H model is expected to be more effective when the data recorded in the previous section is of higher quality and volume. In other words, the performance of the UTTree-H model depends on the number of concepts extracted from patients' histories. EMRs do not behave the same way regarding details recorded in a patient's history. We perform another experiment to investigate conditions that improve the results. A decision tree is made, in which a target field is added for each patient that contains a value of 1 if the UTTree-H method is the best and a value of 0 otherwise. The number of extracted concepts is used as a predictor. The outcomes illustrated in Fig. 6 reveal that UTTree-H results are best when more than eight concepts are detected and extracted from the history section.

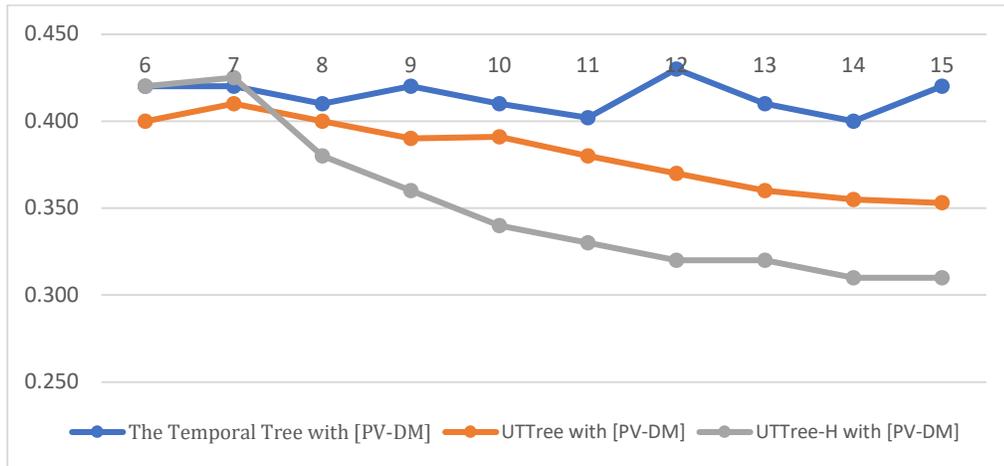

Fig. 6. The effect of the number of extracted words on the error rate

A non-parametric statistical test, called The Wilcoxon test (35), is used to evaluate the differences between the UTTree method and other approaches. The result indicates statistical significance with a p-value of less than 0.01, except the one indicated by an asterisk. For analysis of the selection of hyperparameters, please refer to the supplementary file.

### 4.5 Evaluation metrics in mortality prediction

The effectiveness of embedding vectors is tested on a supervised mortality prediction task. This task is concerned with predicting whether a patient will pass away in the hospital or after discharge. This problem could be formulated as a binary one. The algorithm for the proposed model is given below:

1. The PCA (39) algorithm is utilized to reduce the dimensionality of each dataset's embedding vectors.
2. Random subsets of each dataset are created for training and testing. The proportion of the test split is 0.3
3. The reduced dataset is used to train XGBoost (40), SVM (41), and the random forest (42) classifiers.
4. The model's performance is evaluated considering the accuracy of the test data.
5. Ten-fold cross-validation is used to evaluate the models.

6. The box plot method is employed to identify hidden patterns for visual summarization and comparison of the data groups. Fig. 8 shows the results.

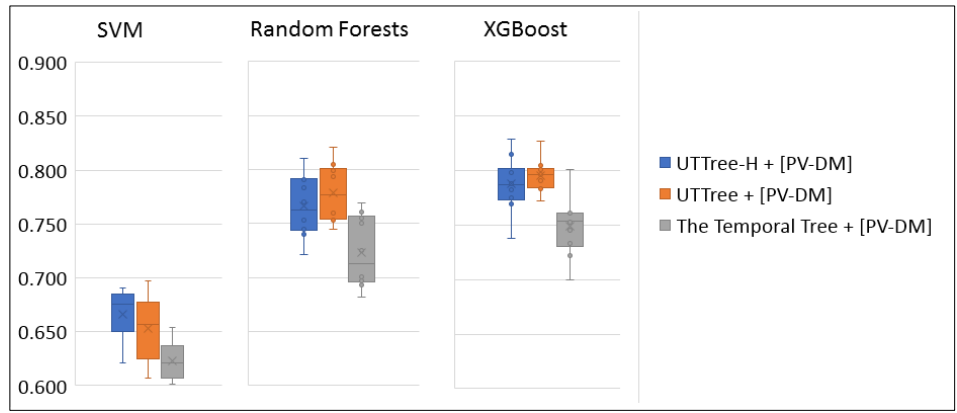

Fig. 7. Predicting mortality in MIMIC-III patients

According to Fig. 7, it is evident that the XGBoost algorithm outperforms RF and SVM in all three datasets. The medians of the proposed model's boxplots are all at the same level. However, the box plots show different distributions of views. It is observed that the box plot of UTTree_ [PV-DM] is comparatively short compared to that of UTTree-H_ [PV-DM] in XGBoost, suggesting that the overall results of UTTree_ [PV-DM] are close to each other. The box plot of UTTree_ [PV-DM] is higher than others.

## 5   Conclusion and future work

This paper applied two data models (UTTree nad UTTree-H) for representing EMR. Using tree structure helped to integrate unstructured and structured data. We preprocessed the unstructured data of EMR, including the clinical narratives, to extract concepts and map them to the UMLS knowledge base.

The experimental results showed that using our approach to extract meaningful information from unstructured data could increase the accuracy of the patient similarity assessment model. We enriched the model by applying the patients' past medical history by proposing the UTTree-H model. We showed under what conditions this method gives better results.

This study suggested that upgrading EMR software and facilitating the registration of information for service providers improve the secondary use of EMR, including patient similarity assessment. The free publicly available hospital database in this paper contained English clinical notes. Since medical records can be generated in languages other than English in many countries, developing the NER system for these languages will be valuable.

The research has several directions for further development. For instance, the weights of different clinical events that make compounds are considered the same in this study. However, in reality, the importance of clinical events would be very different. For example, detecting "fever" and "cancer" in an EMR might have different significance. In future works, the issue of weighting different types of events or different concepts will be examined (24).

Moreover, the knowledge resources for enriching the model structure will be considered. The document-representation algorithms used in this study are context-agnostic methods. Testing the model using context-aware methods, such as transformer-based models, can be studied in another work. Furthermore, we plan to generate different forms of noise in the input text and determine which procedure gives a more robust tree. Another idea is to construct a graph instead of a tree structure and extend the relationships between nodes derived from knowledge sources.

DATA AVAILABILITY

Access to MIMIC-III can be requested at https://physionet.org/content/mimiciii/1.4/.

# Appendix A

A.1 Tools and libraries

Table 5. List of tools and libraries

| Processing step | Tools/python libraries | Setting | Ref |
|---|---|---|---|
| Section detection | medspaCy | nlp.pipe_name : 'medspacy_sectionizer' | (28) |
| Negation detection | Negspacy | | (43) |
| Concept mention detection, Concept encoding | MetaMap | the (2018AA) release of UMLS with the following list of options -V USAbase-L 18 -Z 18 -E -AsI+ –XMLf –negex -E. | (29) |
| | cTAKES | the Default Clinical Pipeline of cTAKES and Python library ctakes-parser 0.1.0[3] | (44) |
| | scispaCy[4] | a Python library with en_ner_bc5cdr_md model | (45) |
| Creating Tree | NetworkX | Python library[5] | (46) |
| Doc2Vec Embedding | Gensim | vector_size = 200, window_size= 5, sampling_threshold = 1e-5, negative_size = 5, alpha = 0.025, minimum_alpha = 0.0001, min_count =5, training_epochs = 10, DM=0 and1,hs=1(for hierarchical softmax) and negative =1 (for negative sampling) | (47) |

A.2 Distance metrics

We examine the cosine and Euclidean metrics to determine the distance between the two patient representation vectors. The result shows that the cosine metric is better than Euclidean one, so this metric is used in all experiments.

Cosine distance: The cosine distance is the measure of similarity between two vectors of an inner product space. Using the cosine of the angle between two vectors, it determines whether two vectors point in roughly the same direction. This distance metric is often used to measure document similarity in text analysis (48):

$$\cos\theta = \frac{v.u}{\|v\|.\|u\|} \tag{6}$$

---

[3] https://pypi.org/project/ctakes-parser

[4] http://allenai.github.io/scispacy

[5] https://networkx.org/

In the equation above, $v$ $and$ $u$ are the TFIDF vectors, and $\theta$ is the angle between two vectors. As θ ranges from 0° to 90°, cos(θ) ranges from 1 to 0.

Euclidean distance: The similarity between two real-valued vectors is computed by the Euclidean similarity. Euclidean distance is calculated as the square root of the sum of the squared differences between the two vectors:

$$d(\vec{v}, \vec{u}) = \|\vec{v} - \vec{u}\| = \sqrt{(v_1 - u_1)^2 + (v_2 - u_2)^2 + \cdots + (v_n - u_n)^2} \tag{7}$$

Table 6. Comparison of cosine and Euclidean distances

|  | TFIDF_S_Cosine | TFIDF_S_Euclidean |
| --- | --- | --- |
| MSE@1 | 0.320 | 0.329 |
| MSE@5 | 0.129 | 0.131 |
| MSE@10 | 0.080 | 0.085 |
| MSE@20 | 0.052 | 0.057 |
| DCG@1 | 0.410 | 0.408 |
| DCG@5 | 0.408 | 0.407 |
| DCG@10 | 0.407 | 0.406 |
| DCG@20 | 0.407 | 0.406 |
| Precision@1 | 0.554 | 0.554 |
| Precision@5 | 0.551 | 0.550 |
| Precision@10 | 0.542 | 0.541 |
| Precision@20 | 0.541 | 0.540 |

A.3 Hyperparameters

Some hyperparameters are tested in this study, including Doc2Vec mode, improvement method, and vector size. The vector size and distance metric in this experiment are fixed at 200 and cosine, respectively. The figure shows that the generated result by PV-DM with hierarchical softmax leads to lower mean error and higher mean nDCG compared to negative sampling. The discussion section covers the possible causes of these changes.

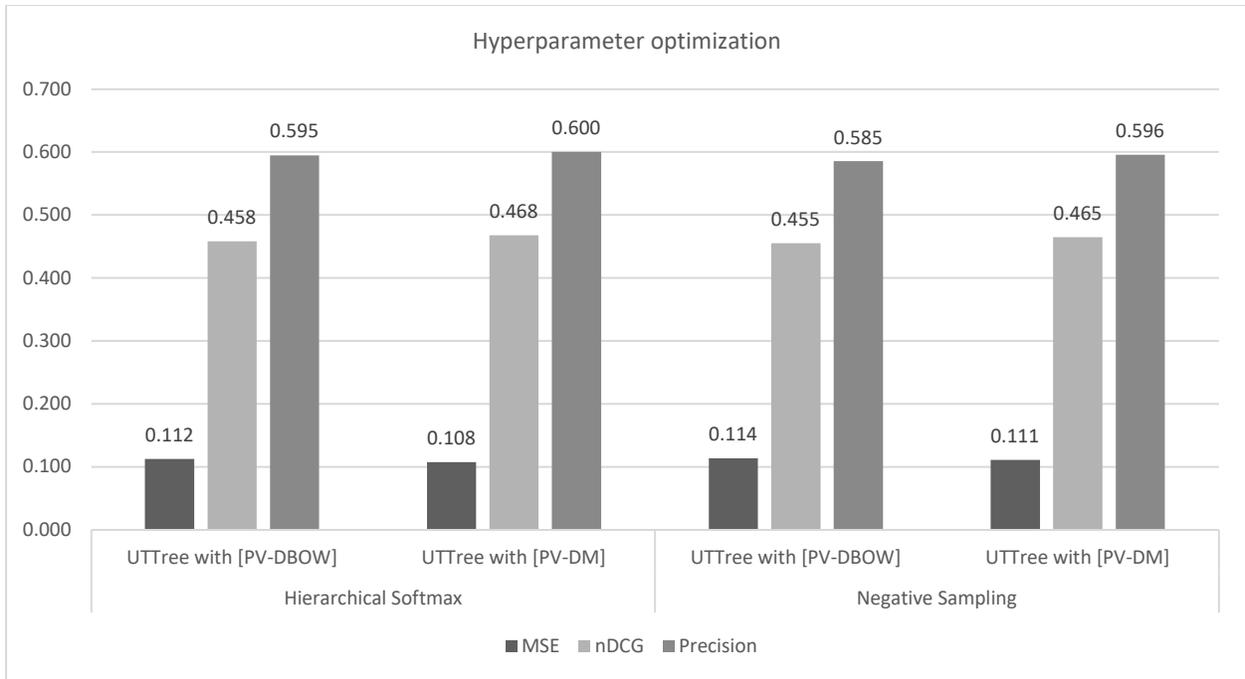

Fig. 8. Comparison of hierarchical softmax and negative sampling

PV-DM performs better than PV-DBOW, but PV-DM takes longer to train. Compared to the negative sampling method, the hierarchical softmax performs better (lower MSE and higher nDCG and precision). The vectors generated by the hierarchical softmax are determined by traversing the network layers rather than evaluating/updating $O(\log(N))O(\log(N))$ network units. The weights are essentially expanded to support a large vocabulary; therefore, hierarchical softmax is better for infrequent words, while negative sampling is better for frequent words (48).

A.4  Embedding vector size

To select the size of the embedding vector, we execute Doc2Vec by hierarchical softmax optimization. Afterward, we test the different vector sizes. The results show a reduction in the MSE by increasing the embedded vector size to 200 properties; therefore, we select vector size 200 for all experiments.

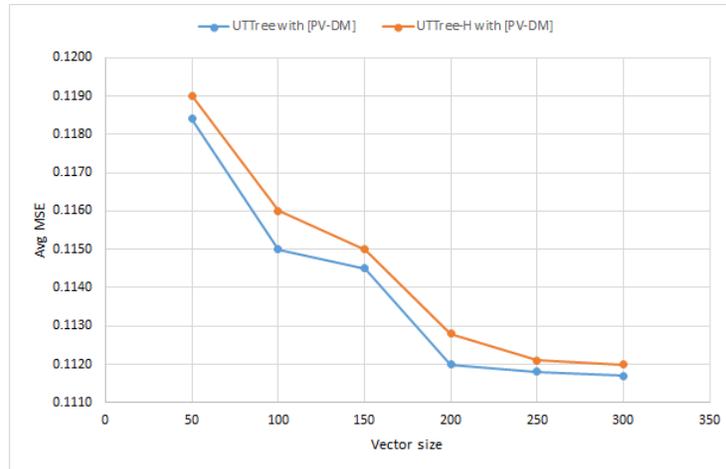

Fig. 8. Embedding vector size

A.5 Pseudo-code of creating tree

*ALGORITHM*

| | |
|---|---|
| 1 | **Input**: $Q_p - $ Repository of quadraples for a $Patient_{ID}$ |
| 2 | //each record in Q has four fields: Time (t), Event − Type (y), Event (e), and Value (v) |
| 3 | **Output**: a string generated by traversing tree |
| 4 | T ← DiGraph() //level 0 |
| 5 | Wp ← Number of Time window for each Patient_ID |
| 6 | α ← Filter $Q_p$ by Temporal Event role = Retrospective data (a) |
| 7 | Foreach w in Wp: //Filter Q by Patient_ID |
| 8 |     add subtree roots |
| 9 |     Foreach s in subtrees: |
| 10 |         Add three nodes for each event type (R: Retrospective data) , (N: New founded Data), (SH: Short effect) |
| 11 |         Add α to R as new nodes |
| 12 |         β ← Filter $Q_p$ by Temporal Event role: New founded Data & w |
| 13 |         γ ← Filter $Q_p$ by Temporal Event role: Short effect & w |
| 14 |         Add β to N as new nodes |
| 15 |         Add γ to SH as new nodes |
| 16 | Foreach n in non − leaf nodes: |
| 17 |     Update label(n) ← concat label of nodes in next level |
| 18 | Breadth-First traversal T |